# Bridging Language Gaps: Utilizing Interactive Robots to Teach Cantonese in Real-Life Contexts for Newly-Arrived Children

KA-YAN FUNG, YUXING TAO, TZE-LEUNG, RICK LUI, and KUEN-FUNG SIN, The Education Univeristy of Hong Kong, China

Hong Kong's education system is notably multicultural, including local, non-Chinese-speaking, and newly arrived students (NAS) (Mandarine Chinese-speaking). NAS can guess the meaning of vocabulary but cannot speak out, presenting unique challenges for them, particularly language barriers and cultural differences. These challenges hinder their academic success and social integration, leading to feelings of isolation and demotivation. Current resources often fail to address the emotional well-being of these students and predominantly focus on English language acquisition, leaving a gap in support for learning Cantonese and navigating the local cultural landscape. This study explores the effectiveness of an interactive robot, *Boon Boon*, in teaching Cantonese through real-life contexts to enhance NAS children learning engagement and motivation. The research questions are: (1) How does interactive robot-empowered scenario learning influence the learning engagement and motivation of NAS in learning Cantonese? and (2) What is the impact of a robot-empowered scenario learning system on the Cantonese language proficiency of NAS? Fourteen children are invited to participate in a four-day learning program with *Boon Boon*. The preliminary result indicated that *Boon Boon* drove students' attention to learning and academic achievement. Future research will focus on long-term assessments of robot-empowered learning's effectiveness and explore the scalability of this approach across diverse educational settings and cultural backgrounds.

## 1 Introduction

Hong Kong's education system is characterized by its diverse and multicultural environment, encompassing students with and without special learning disabilities (SpLDs), non-Chinese speaking (NCS) students and newly-arrived students (NAS) (Mandarin Chinese-speaking). NAS often face hurdles in adapting to the local educational landscape, primarily due to language barriers, cultural differences, and social integration issues [6]. In recent years, the integration of robot-assisted language learning (RALL) has shown strengthened educational commitment. Interactive robots are increasingly utilized in primary schools to capture students' attention and enhance their learning proficiency, particularly in inclusive learning environments [5]. Also, robot-empowered systems can facilitate real-life scenario practice, helping students navigate social norms, dining etiquette, and public behaviour.

Authors' Contact Information: Ka-Yan Fung, fkayan@eduhk.hk; Yuxing Tao; Tze-Leung, Rick Lui; Kuen-Fung Sin, The Education Univeristy of Hong Kong, Hong Kong, China.

### 1.1 Language Acquisition Challenges for NAS

NAS encounter many challenges in language acquisition that impact their academic and social integration, particularly in diverse educational environments like Hong Kong. A primary hurdle is the language gap, where NAS often struggle to communicate in Cantonese and recognize Traditional Chinese effectively [9], leading to increased cognitive burden and mental pressure [16]. This condition creates substantial barriers to their effective communication and learning.

### 1.2 Interactive Learning and Robot-empowered Technology in Education

Interactive learning with robot-empowered technology in education involves utilizing robotic systems to create engaging and student-centered experiences. This approach, particularly through the use of interactive robots, such as *Boon Boon*, ultimately increases learning inspiration and skill acquisition through reciprocal interaction [2]. Previous studies have demonstrated the effectiveness of *Boon Boon* in educational settings, showcasing its ability to foster learning engagement among diverse student populations.

### 1.3 SDT in Education

SDT emphasizes that human motivation flourishes when three psychological needs (i.e., autonomy, competence, and relatedness) are met [12]. SDT has been widely applied in educational technology and language learning to facilitate student engagement and learning performance [3]. In educational technology, SDT principles inform the design and evaluation of digital systems that promote intrinsic motivation [10]. For instance, customizable learning platforms support autonomy, while gamified applications provide feedback that fosters a sense of competence [11].

### 1.4 Real-Life Contexts in Language Learning

Real-life contexts in language learning are grounded in the principles of communicative language teaching (CLT) [1], which emphasizes the importance of using language in authentic scenarios to magnify communicative competence and cultural understanding [15]. Language acquisition is effective when tied to practical, relatable situations rather than abstract drills [17]. By engaging learners in meaningful interactions that mirror everyday situations, educators can help students refine their ability to transmit and interpret messages effectively. Thus, real-life simulation can foster linguistic proficiency and sociocultural awareness. Despite the availability of different support tools [13].

To address these gaps, we propose an interactive robot, *Boon Boon*, to teach Cantonese in real-life contexts for NAS to build up their learning incentive through the lens of self-determination theory (SDT). This study aims to explore the effectiveness of *Boon Boon* in revamping learning interests and efficacy for NAS by investigating the following research questions (RQs): *RQ1) How does interactive robot-empowered scenario learning influence the learning engagement and motivation of NAS in learning Cantonese?* and *RQ2) What is the impact of a robot-empowered scenario learning system on the Cantonese language proficiency of NAS?* To respond to the two RQs, 14 newly-arrived children were recruited to participate in a four-day learning programme.

## 2 Scenario System Design

In this work, we designed a scenario learning system incorporating with real-life contexts on an interactive robot, *Boon Boon* [8]. The real-life scenarios and user interfaces (UIs) were co-designed by two professional specializing in language learning and human-robot interaction (HRI).

*Interactive Robots.* *Boon Boon* has one head with a 9-inch HD display, a pair of soft ears, two pairs of auxiliary and swivel wheels. The soft ears can act as pressure relief functions. *Boon Boon* can also "walk and dance" through rolling wheels, allowing an interactive learning.

*Real-life Scenario Content.* The scenario content was co-designed with an education professional and reviewed by an experienced teacher. The system consists of three scenarios covering supermarket, restaurant, and transportation.

*Set 1: Supermarket* aims to teach NAS how to behave appropriately in a supermarket setting by focusing on shopping carts and baskets. The objectives: (1) to engage students in understanding the purpose of these tools and assessing their recognition of their functions, (2) the curriculum emphasizes safety awareness and guides students on how to use shopping carts and baskets. This enables NAS to learn the social norms and expectations within the local context.

*Set 2: Restaurant* aims to teach NAS the appropriate behaviours and etiquette when visiting a restaurant. The first objective is to engage students in recognizing the procedures involved in entering a restaurant. Next, the curriculum focuses on teaching students how to utilize utensils properly. Additionally, students learn about the etiquette expected in a restaurant, which is essential for social integration in the local culture.

*Set 3: Transportation* aims to teach NAS about railways and the appropriate behaviours associated with using them. The first objective is to allow students to understand the procedure of entering the railway station. Following this, students learn the proper steps for boarding a train, essential for their confidence and safety when travelling.

*Interface Design.* The UIs of the system are designed according to a standardized specification, which ensures consistency and usability across various real-life scenarios. The elements, such as font size, background colour, and functional buttons, follow the standardized grid to design. The aim is to ensure consistency and provide high colour contrast, ultimately enhancing readability and accessibility. Moreover, the robot would respond to the students whether their answers were correct.

*System Design.* The scenario learning system was developed using Android Studio and integrated automatic speech recognition (ASR) technology, featuring both Speech-to-Text (STT) and Text-to-Speech (TTS) functionalities. This system enabled the conversion of spoken language into text and supports transcription and translation in Traditional Chinese (Cantonese). The system provided instant feedback by comparing student input with the pre-programmed transcription and translation, helping students enhance their language skills.

## 3 Experiment

This section elaborates on the experiments designed to investigate the efficacy of the interactive robot-empowered scenario learning system and answer the RQs.

*Participants.* 10 students (5 males and 5 females) aged 6 to 10 from a local primary school ($\bar{M}$= 8.33-year-old, $SD$=0.83) participated in the study. The inclusion criteria for students to join the study were: (1) studying in Grades 1 to 3; (2) having at least one year of experience in learning Traditional Chinese and speaking Cantonese; (3) being NAS; (4) having no other medical or physical disabilities that might affect their interaction with the robots, cognitive and reading ability; and (5) having experience in utilizing digital tools, such as tablets. Before the experiment, we obtained informed consent from the students' guardians. Participation was entirely voluntary and contingent upon their consent. The experimental protocol received approval from the University Institutional Review Board (IRB). Participants were not offered any form of remuneration.

*Questionnaires.* The pre- and post-questionnaires consisted of four variables (each variable includes three sub-question) categorized into motivation and engagement. The questionnaires were utilized in the same region [4], and we modified them based on our study. An experienced teacher reviewed the items to ensure the clarity of the questionnaire. The questionnaire consists of pre- / post- engagement and motivation, including *Behavioural Engagement* (e.g., I try my best to finish all learning activities), *Emotional Engagement* (e.g., I feel engaged when learning with the robot), *Cognitive Engagement* (e.g., I review my work to ensure its accuracy), *Motivation* (e.g., I find joy in learning scenarios).

## 4 Data Analysis

Data analysis is conducted to measure and answer the two research questions, including (1) NAS learning engagement and motivation and (2) learning performance.

*RQ1: How does interactive robot-empowered scenario learning influence the learning engagement and motivation of NAS in learning Cantonese?* RQ1 assesses the extent to which interactive robot-empowered scenario learning influences the learning engagement and motivation of NAS in learning Cantonese. The analysis of paired-samples t-test result shows that learning with interactive robots can promote students' engagement in the behaviour, emotional, cognitive and intrinsic motivation.

Behavioural Engagement. NAS students improved by 13.56%, $p < .01$ (pre-test: $\bar{M} = 78.67$, $\sigma = 14.33$; post-test: $\bar{M} = 89.33$, $\sigma = 8.43$).

Emotional Engagement. NAS students improved by 4.76%, $p = .05$ (pre-test: $\bar{M} = 84.00$, $\sigma = 11.42$; post-test: $\bar{M} = 88.00$, $\sigma = 8.78$).

Cognitive Engagement. NAS students improved by 3.25%, $p = .46$ (pre-test: $\bar{M} = 82.00$, $\sigma = 12.59$; post-test: $\bar{M} = 84.67$, $\sigma = 11.78$).

Intrinsic Motivation. NAS students improved by 0.74%, $p = .76$ (pre-test: $\bar{M} = 90.00$, $\sigma = 11.00$; post-test: $\bar{M} = 90.67$, $\sigma = 10.98$).

From the descriptive statistics mentioned above, we found that NAS improved in three engagement and motivation after a four-day learning. Behavioural, emotional engagement, and motivation performance variations were less dispersed.

*RQ2: What is the impact of a robot-empowered scenario learning system on the Cantonese language proficiency of NAS?.* RQ2 evaluates the impact of a robot-empowered scenario learning system on the Cantonese language proficiency of NAS. The result shown that NAS students improved by 5.41%, $p = .83$, $f = .05$ (pre-test: $\bar{M} = 61.67$, $\sigma = 33.38$; post-test: $\bar{M} = 65.00$, $\sigma = 33.75$).

## 5 Discussion

The interactive features and positive reinforcement provided by *Boon Boon* foster behavioural engagement and enhance emotional engagement among NAS students, creating a supportive and effective learning environment. We discuss how *Boon Boon* impacts NAS learning engagement, motivation, and language proficiency. We also explore the need for comprehensive support and the potential for broader application.

*Learning Engagement.* Instant feedback and interactive responses are crucial in promoting students' active participation. Our results demonstrated that *Boon Boon* can escalate students' learning interests, with evidence in the substantial increase in behavioural engagement (+6.94%). Positive reinforcement can encourage students to actively engage in the learning process, which is aligned with previous research that underscores the importance of feedback in fostering language acquisition [5, 7].

*Language Proficiency.* The scenario-based training follows the context-driven learning curriculum that can cultivate students' understanding. In our study, NAS improved Cantonese language proficiency by 5.66%. The scenario-based training can streamline authentic vocabulary acquisition, allowing them to visualize and articulate their understanding of a new language [14].

## 5.1 Limitations

We would like to acknowledge several limitations in this study. First, our study involved a relatively small sample size as a preliminary investigation. A larger sample size would provide more robust insights, but it would not limit our findings' generalizability. Second, due to constraints in the school schedule, we conducted a four-day study, which may be insufficient for capturing the substantial effects of RALL. A long-term study allows for a more comprehensive assessment of the sustained impact of this educational approach. In the future, we will conduct a comparative study to further consolidate the effect of RALL.

## 6 Conclusion

This work demonstrated that RALL positively aroused the enthusiasm of NAS students to learn actively. After four days of learning, NAS enhanced behaviour, emotion and cognition, ultimately creating a happy learning environment. The short-term study has provided some design opportunities for RALL in NAS academic success. For example, students changed from shy to talkative after interacting with *Boon Boon*, and they learned with robots without peer pressure. If *Boon Boon* can adapt to NAS individual needs, such as detecting and correcting their accent, the system can deepen NAS learning efficacy. The system can further contribute to the community with a larger scale, longer-term and broader beneficial groups (e.g., ADHD, NCS, and older adults) with other languages, such as English and Japanese.

## 7 Proposal Outline

(1) Title: Bridging Language Gaps: Utilizing Interactive Robots to Teach Cantonese in Real-Life Contexts for Newly-Arrived Children
(2) Context of the Demo: To demonstrate interactions, functions and learning process of the robot-empowered learning system
(3) Duration: 30-minute
(4) Required Equipment: Wi-Fi and a Type-C USB-supported socket
(5) Required Space: A 1.5 square-feet table
(6) Installation: To install the apk in *Boon Boon*
(7) Contribution to HCI:
   - Current Language Learning Challenges: NAS face language barriers, cultural differences, and social integration issues. Existing resources focus more on English acquisition, neglecting Cantonese and local culture. NAS students struggle with the transition from Mandarin to Cantonese.
   - Interactive Robots in Education: RALL is effective in engaging students and enhancing learning. Interactive robots, such as *Boon Boon* capture attention and improve learning proficiency. *Boon Boon* offers real-time feedback and adaptive learning strategies, encouraging active learning.

- Scenario-based Learning: Real-life contexts enhance language learning by providing authentic scenarios. The scenario learning system includes three sets: supermarket, restaurant, and transportation. Each set is designed to teach appropriate behaviours and etiquette in specific settings.
- Rationale: To address the gap in comprehensive language support programs for NAS. The system can foster learning engagement and motivation through interactive and scenario-based learning. Through the utilization of SDT, the system can enhance students' intrinsic motivation and engagement.
- Novelty: To explore the use of interactive robots in teaching Cantonese to NAS in real-life contexts. This study shows positive impact on NAS students' engagement, motivation, and language proficiency. The result highlights the potential for broader application of RALL in diverse educational settings.